\documentclass[times,twocolumn]{aastex631}

\usepackage{CJK}
\usepackage{amsmath, amssymb, bm}
\usepackage{mathrsfs}
\usepackage{color}
\usepackage[nodayofweek]{datetime}
\newdateformat{monthyearday}{%
  \THEYEAR\ \monthname[\THEMONTH] \twodigit{\THEDAY}}
\usepackage{multirow}

\makeatletter
\patchcmd{\NAT@citex}
  {\@citea\NAT@hyper@{%
     \NAT@nmfmt{\NAT@nm}%
     \hyper@natlinkbreak{\NAT@aysep\NAT@spacechar}{\@citeb\@extra@b@citeb}%
     \NAT@date}}
  {\@citea\NAT@nmfmt{\NAT@nm}%
   \NAT@aysep\NAT@spacechar\NAT@hyper@{\NAT@date}}{}{}

\patchcmd{\NAT@citex}
  {\@citea\NAT@hyper@{%
     \NAT@nmfmt{\NAT@nm}%
     \hyper@natlinkbreak{\NAT@spacechar\NAT@@open\if*#1*\else#1\NAT@spacechar\fi}%
       {\@citeb\@extra@b@citeb}%
     \NAT@date}}
  {\@citea\NAT@nmfmt{\NAT@nm}%
   \NAT@spacechar\NAT@@open\if*#1*\else#1\NAT@spacechar\fi\NAT@hyper@{\NAT@date}}
  {}{}
\makeatother

\newcommand{\uat}[2]{\href{http://astrothesaurus.org/uat/#1}{#2  (#1)}}
\newcommand{\hi}{\textrm{H\,{\sc i}}}
\newcommand{\nhi}{{N_{\textrm{H\,{\sc i}}}}}
\newcommand{\mhi}{{M_{\textrm{H\,{\sc i}}}}}

\newcommand{\s}{{s$^{-1}$}}
\newcommand{\kms}{km\,s$^{-1}$}
\newcommand{\mjb}{mJy\,beam$^{-1}$}
\acceptjournal{The Astrophysical Journal Letters}

\shorttitle{Mpc-Scale \hi\ Flows in the Neighborhood of HCG 100}
\shortauthors{Yu et al.}
\graphicspath{{./}{figures/}}

\begin{document}
\begin{CJK*}{UTF8}{gbsn}
\title{Megaparsec-Scale Neutral Hydrogen Flows in the Neighborhood of Hickson Compact Group 100 }

\author[0000-0003-3230-3981]{Qingzheng Yu (余清正)}
\affiliation{Dipartimento di Fisica e Astronomia, Universit\`a degli Studi di Firenze, Via G. Sansone 1, 50019 Sesto Fiorentino, Firenze, Italy; \url{qingzheng.yu@unifi.it}}
\affiliation{INAF - Osservatorio Astrofisico di Arcetri, Largo E. Fermi 5, I-50125 Firenze, Italy}

\author[0000-0002-2853-3808]{Taotao Fang (方陶陶)}
\affiliation{Department of Astronomy, Xiamen University, Xiamen 361005, People's Republic of China; \url{fangt@xmu.edu.cn}}

\author[0000-0003-4019-0673]{Enrico M.\ Di Teodoro}
\affiliation{Dipartimento di Fisica e Astronomia, Universit\`a degli Studi di Firenze, Via G. Sansone 1, 50019 Sesto Fiorentino, Firenze, Italy; \url{qingzheng.yu@unifi.it}}
\affiliation{INAF - Osservatorio Astrofisico di Arcetri, Largo E. Fermi 5, I-50125 Firenze, Italy}

\author[0000-0003-0202-0534]{Cheng Cheng (程诚)}
\affiliation{Chinese Academy of Sciences South America Center for Astronomy, National Astronomical Observatories, CAS, Beijing 100101, People's Republic of China}
\affiliation{National Astronomical Observatories, Chinese Academy of Sciences, 20A Datun Road, Chaoyang District, Beijing 100101, People's Republic of China}

\author[0000-0002-1588-6700]{Cong Kevin Xu (徐聪)}
\affiliation{Chinese Academy of Sciences South America Center for Astronomy, National Astronomical Observatories, CAS, Beijing 100101, People's Republic of China}
\affiliation{National Astronomical Observatories, Chinese Academy of Sciences, 20A Datun Road, Chaoyang District, Beijing 100101, People's Republic of China}



\begin{abstract}
The evolution of galaxies is strongly influenced by their ability to exchange gas with their surroundings, yet direct observational constraints on these processes remain scarce. 
Using ultra-deep neutral hydrogen (\hi) observations with the Five-hundred-meter Aperture Spherical Telescope (FAST), we detect a diffuse \hi\ structure extending over $\sim$1 Mpc around the compact galaxy group HCG 100, with integrated column densities down to $\sim$$8.6\times10^{17}$ cm$^{-2}$.
This structure is among the most extended and lowest-density neutral gas systems ever observed in emission.
The \hi\ gas forms a coherent envelope connecting the compact group to neighboring galaxies and shows a large-scale coherent velocity gradient around HCG 100 extending across $\sim$0.6 Mpc.
The extended structure contains $\sim$$1.4\times10^{10}$ $M_\odot$ of diffuse neutral gas, implying that up to $\sim$40$-$50\% of the \hi\ gas resides outside galaxies. Such a large diffuse \hi\ structure may arise either from large-scale tidal debris or gas accretion from the surrounding cosmic web.
Our results demonstrate that massive reservoirs of diffuse neutral gas can persist on megaparsec scales around galaxy groups for extended periods, providing a previously unseen component of baryon cycling in dense environments.
\end{abstract}


\section{Introduction} \label{sec:intro}
Cosmological simulations predict that galaxy evolution is regulated by the continuous exchange of gas between galaxies and their surrounding environment. Gas may accrete onto galaxies from the cosmic web through both ``cold'' and ``hot'' modes, i.e., either streaming along filaments as relatively cool gas or cooling from shock-heated halos \citep{Kere2005,Dekel2006,Dekel2009}, thereby replenishing their gas reservoir and sustaining star formation. Interactions and mergers can redistribute baryons over large spatial scales \citep{Mihos1996,Hopkins2008}, creating very complex environments. 
Atomic neutral hydrogen (\hi) is a powerful tracer of these processes, being highly sensitive to tidal interactions, ram-pressure effects, and accretion. Observations of \hi\ therefore provide a unique window into gas flows across galactic and intergalactic environments over long dynamical timescales \citep[e.g.,][]{Yun1994,Zhu2021,Xu2022,Lawrie2025}.

Hickson Compact Group 100 (HCG 100, distance $D=77\pm5$ Mpc) is a system of four late-type galaxies (HCG 100a, HCG 100b, HCG 100c, HCG 100d) embedded in a dynamically active environment \citep{Hickson1982,Hickson1992}. Previous studies have revealed clear signatures of strong multi-galaxy interactions \citep{Plana2003,Jones2023}, including extended \hi\ tidal tails \citep{Jones2023,deMello2008}, star-forming regions in the intragroup medium \citep{deMello2008,Vilchez1998,Torres-Flores2009}, and candidate tidal dwarf galaxies \citep{Torres-Flores2009,deMello2012}. These features indicate ongoing baryon redistribution within the group and make HCG 100 an ideal laboratory for studying the cycling of gas in dense galaxy environments.

To probe the neutral gas content around HCG 100, we carried out very deep mapping of the \hi\ 21-cm emission in the group and its surroundings using the Five-hundred-meter Aperture Spherical radio Telescope (FAST). Besides the four member galaxies within HCG 100, nearby galaxies MRK 935, NGC 7810, and AGC 105092 outside the group are also covered in the mapping region. Our new data reach a \hi\ column-density sensitivity of $4.8\times10^{16}\,\rm cm^{-2}$ per 20 ${\rm km\,s^{-1}}$ channel, with an angular resolution of 4$^{\prime}$, corresponding to $\sim90$ kpc at the distance of HCG 100. 
At this sensitivity, the observations reveal a previously undetected, large-scale diffuse \hi\ structure that links HCG 100 with its neighboring galaxies.

The remainder of this Letter is structured as follow. After describing our new FAST data in Section~\ref{sec:obs}, we present the results on the diffuse \hi\ in and around the group in Section~\ref{sec:results}. We discuss our findings in Section~\ref{sec:discussions} and conclude in Section~\ref{sec:Summary}.
Throughout this work, we assume a standard $\Lambda$ Cold Dark Matter ($\Lambda$CDM) cosmology with H${_0}=70$ km s$^{-1}$ Mpc$^{-1}$, $\Omega_{\Lambda}=0.7$, and $\Omega_\mathrm{m}=0.3$.

\section{Observations and Data Reduction} \label{sec:obs}
\subsection{FAST Observations}
Deep \hi\ mapping observations were conducted in two cycles from September 2023 to October 2024 using the FAST 19-beam receiver in standard ON-OFF mode, with a total observing time of 31.8 hours including overheads. The FAST 19-beam L-band receiver consists of 19 beams arranged in a hexagonal configuration with a spacing of 5$\farcm$7 between adjacent beams \citep{Jiang2020}. The observations were centered at 1395.56 MHz, with a frequency coverage of 1050$-$1450 MHz and a spectral resolution of 7.63 kHz ($\Delta v = 1.6\ {\rm km\,s^{-1}}$). The average half-power beamwidth at 1395.56 MHz is $\sim$2$\farcm$9 across the 19 beams. 

\begin{figure*}[!htbp]
\centering
\includegraphics[width=1.0\textwidth]{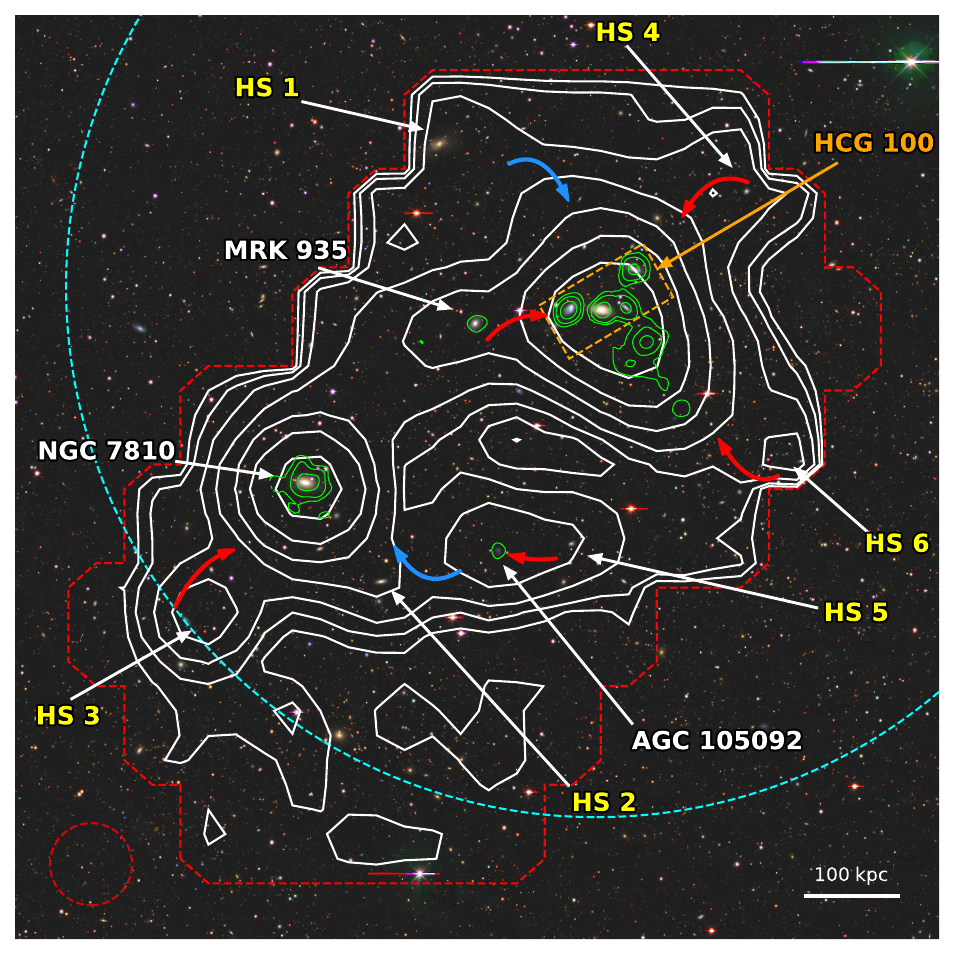}
\caption{
The \hi\ gas in the velocity range of 5000$-$5700 ${\bf km\,s^{-1}}$. The contour map of the integrated \hi\ emission in the velocity range of 5000$-$5700 ${\rm km\,s^{-1}}$ is overlaid on the DECaLS composite color image. The white contours from FAST observations start from $\nhi=8.6\times10^{17}\,\rm cm^{-2}$ (3$\sigma$), with an increment of a factor of two. The green contours show the integrated \hi\ emission between 5000$-$5700 ${\rm km\,s^{-1}}$ adopted from VLA observations (angular resolution of $50^{\prime\prime}\times58^{\prime\prime}$), starting from $\nhi=9.3\times10^{19}\,\rm cm^{-2}$ (3$\sigma$) and increasing by a factor of two. The red dashed circle in the bottom left corner represents the FAST beam after smoothing (4$^{\prime}$). The red dashed lines outline the boundary of the FAST observations, while the cyan dashed circle represents the footprint of the VLA observations. The orange dashed rectangle illustrates member galaxies of HCG 100. All known galaxies and newly detected \hi\ sources (HS in yellow) are marked with white arrows. The red or blue arrows illustrate positive or negative velocity gradients of the \hi\ gas toward nearby converging directions.
\label{fig:hi_flows_overlay}
}
\end{figure*}

In the first observing cycle, 16 pointings were performed in a $4\times4$ rectangular grid for a region of $\sim30^{\prime}\times30^{\prime}$ centered on HCG 100, following the same configuration of \hi\ observations for Stephan's Quintet \citep{Xu2022}. This observing setup meets the Nyquist sampling criterion and ensure complete coverage of the focal plane. Observations for the second field were performed with the same configuration toward the southwest of the previous one. The observation details are summarized in the Table \ref{tab:obs}. The final map covers an area of $\sim30^{\prime}\times57^{\prime}$ centered on HCG 100 and its neighbors \textbf{(Figure \ref{fig:hi_flows_overlay})}, comprising 560 sky pixels (beam positions), with a spacing of 1$\farcm$4 in right ascension and 1$\farcm$2 in declination. The 1$\sigma$ pointing error of individual beams is 7$\farcs$9. At each pointing of observing cycle 1 and 2, six and three pairs of ON-OFF integrations were conducted with the OFF position located 40$^{\prime}$ southeast of the ON position, respectively. For the two observing cycles, each ON or OFF scan lasted 300 s and 260 s, with a sampling rate of 1 Hz. The total on-target integration time per sky pixel is 1800 s and 780 s (Table \ref{tab:obs}), respectively. To minimize the impact of standing waves and sidelobe contamination, all observations were restricted to zenith angles less than 26$^\circ$. 

\subsection{Data Reduction and Calibration}
For each sky pixel observed by a given beam, spectral data were reduced and calibrated following a standard procedure \citep{Xu2022,Liu2024}. The two polarizations were processed separately and subsequently combined after a consistency check. The calibrated spectra (converted from ADU to antenna temperature in K) were averaged to yield a single raw spectrum per integration. During each ON-OFF cycle, a 10 K calibration signal was injected for 20 seconds at the beginning of the cycle. These data were used to calibrate the antenna temperature ($T_{a}$). The calibration error is on the order of 10\%. Each ON or OFF observation thus produced a raw calibrated spectrum, which was then converted from $\rm T_{a}$ (K) to flux density (Jy). The gain factor (in units of K/Jy) used for this conversion is frequency-dependent and varies between beams. For each beam at 1395.56 MHz, the gain was interpolated from frequency-dependent values \citep{Jiang2020,Liu2024}. Radio frequency interference (RFI) was manually flagged, and baseline and standing wave features were removed by fitting a combined sinusoidal and low-order polynomial model. Frequency was converted to velocity using the optical definition of redshift and referenced to the barycentric frame. The spectrum was then re-binned into velocity bins of $\Delta v = 20\ {\rm km\,s^{-1}}$. This entire reduction process was repeated for each sky pixel in the observed field.

A sidelobe correction was applied to the data cube using images of individual beams of the 19-beam receiver \citep{Jiang2020}. The full procedure is described in \ref{app:sidelobe}. The modeled sidelobe contamination is concentrated around the peaks of the bright \hi\ sources in the velocity range of 5200$-$5600 \kms\ (Figure \ref{fig:sidelobe}), with a mean value of 5.5 \mjb\,\kms\ and a maximum of $\sim$ 50 \mjb\,\kms\ near the bright-source peaks, while the diffuse emission between the sources is typically detected at 68$-$136 \mjb\,\kms. The local sidelobe level at the positions of the main diffuse structures is much lower than the detected emission, indicating that the adopted correction does not produce bridge-like residuals at the level of the diffuse \hi\ (Figure \ref{fig:sidelobe}).

The data cube was assembled from 608 individual spectra, with velocity coverage of 3960$-$6740 ${\rm km\,s^{-1}}$ in 20 ${\rm km\,s^{-1}}$ channel bins. We adopted a uniform half-power beam width (HPBW) of 2$\farcm$9  for all spectra, neglecting the slight beam-to-beam variation. The actual HPBW of individual beams at 1395.56 MHz is interpolated from the values in \cite{Jiang2020}. The flux density of the spectra in the cube is given in ${\rm mJy\,beam^{-1}}$. 

Because adjacent beams overlap (beam-separation: 1$\farcm$4$\times$1$\farcm$2), the \hi\ data cube is highly redundant within a single 2$\farcm$9 beam footprint. We exploit this overlap by convolving with a Gaussian kernel (i.e., smoothing) when generating channel and integrated-intensity maps. This smoothing minimizes the noise due to beam-to-beam fluctuations and results in a significant improvement in the \hi\ column density sensitivity. The only cost is a slight degradation in the angular resolution. For an unsmoothed channel map, the mean root mean square (rms) noise of 0.17 ${\rm mJy\,beam^{-1}}$ corresponds to a $1\sigma$ \hi\ column density sensitivity of $1.4\times10^{17}\,\rm cm^{-2}$ per channel (${\Delta v = 20\,\rm km\,s^{-1}}$). The mean rms of smoothed data cube is 0.065 \mjb. Convolution with a Gaussian with full width at half maximum (FWHM) of 2$\farcm$8 improves the sensitivity by a factor of 2.9 to $1\sigma = 4.8\times10^{16}\,\rm cm^{-2}$ per channel, while the angular resolution is degraded only slightly (by a factor 1.4) to 4$^{\prime}$. This increased \hi\ column density sensitivity is crucial for detecting diffuse, extended emission.

\begin{figure*}[!htbp]
\includegraphics[width=0.99\textwidth]{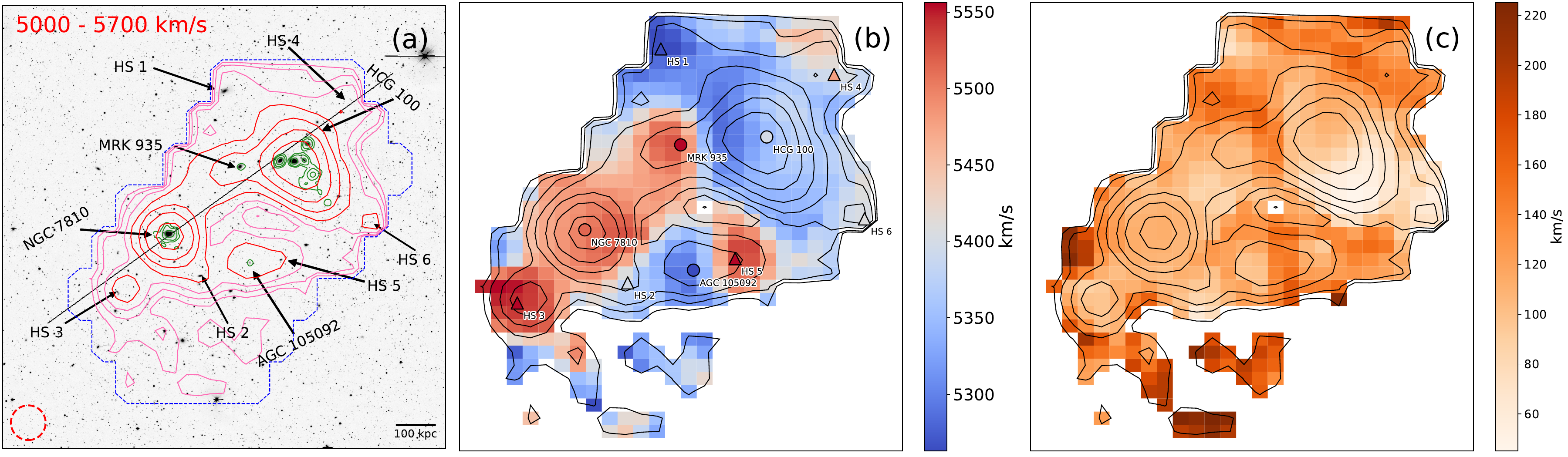}
\includegraphics[width=0.99\textwidth]{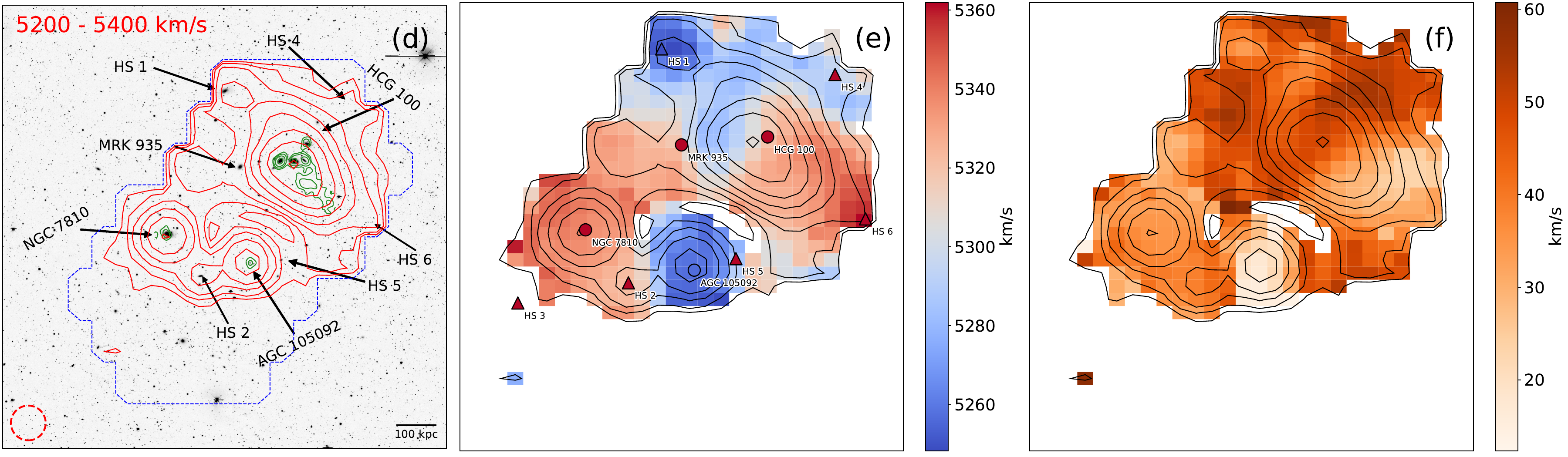}
\includegraphics[width=0.99\textwidth]{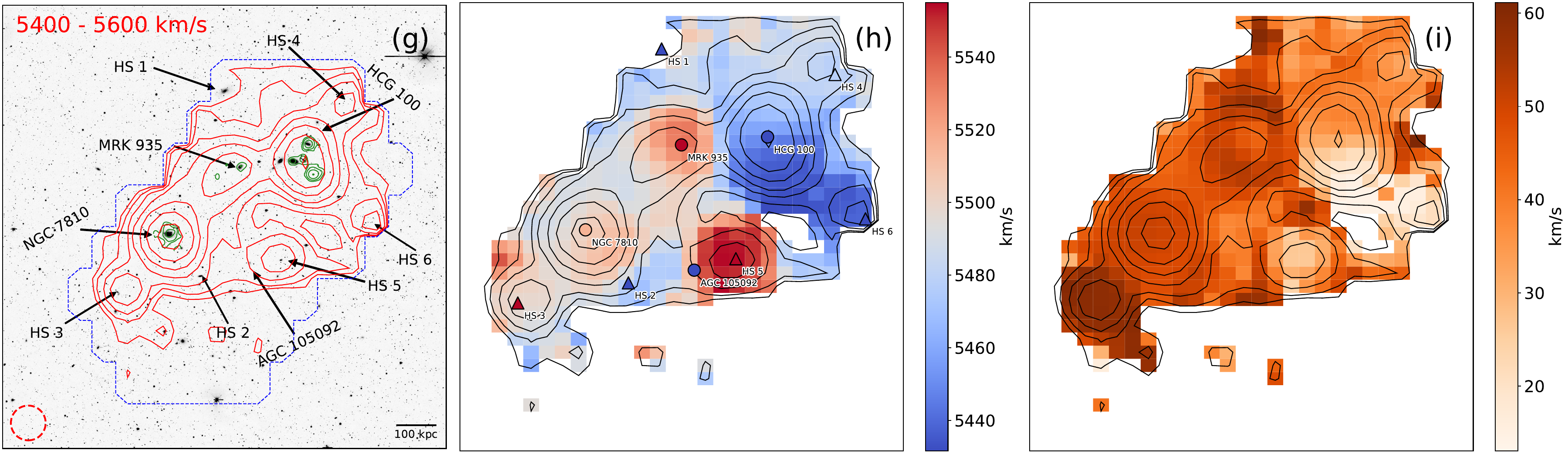}
\caption{
\hi\ gas distribution and kinematics.
Panels (a), (d), and (g): contours of the integrated \hi\ emission in the velocity ranges of 5000$-$5700 ${\rm km\,s^{-1}}$, 5200$-$5400 ${\rm km\,s^{-1}}$, and 5400$-$5600 ${\rm km\,s^{-1}}$ overlaid on the DECaLS $r$-band image. The FAST contours (pink and red) and the VLA contours (green) in panel (a) have the same levels as Figure \ref{fig:hi_flows_overlay}. The pink contours in panel (a) represent very low-density gas with $\nhi<3.4\times10^{18}\,\rm cm^{-2}$ (12$\sigma$).  The black line across the contours illustrates the major axis of the \hi\ structure around HCG 100. In panels (d) and (g), the FAST contours (red) start from $\nhi=4.6\times10^{17}\,\rm cm^{-2}$ (3$\sigma$), while the VLA contours have a starting level of $\nhi=5.0\times10^{19}\,\rm cm^{-2}$ (3$\sigma$). The red dashed circle in the bottom left corner represents the FAST beam after smoothing (4$^{\prime}$). The blue dashed lines outline the boundary of the FAST observations. Panels (b), (e), and (h): velocity field of the gas overlaid with contours from panels (a), (d), and (g). Dots and triangles correspond to the detected galaxies, which are colored by their \hi\ velocities in accordance with the colorbar. Panels (c), (f), and (i): line width maps of the gas overlaid with contours from panels (a), (d), and (g). The velocities and velocity widths of the \hi\ emission are derived from moment maps.
\label{fig:moment_maps}
}
\end{figure*}

\section{Results} \label{sec:results}

\subsection{Diffuse Mpc-scale Neutral Gas Envelope }\label{subsec:hi_envelop}
Figure \ref{fig:hi_flows_overlay} presents the integrated \hi\ emission in the barycentric optical velocity range 5000$-$5700 ${\rm km\,s^{-1}}$ (white contours), overlaid on the DECam Legacy Survey (DECaLS) composite colour image \citep[$g$, $r$, $z$ bands;][]{Dey2019}. 
In contrast to previous \hi\ observations with the Very Large Array \citep[VLA, green contours;][]{deMello2008,Jones2023}, which detected only high-density gas ($\nhi\geqslant9.3\times10^{19}\,\rm cm^{-2}$ over 700 \kms) in HCG 100 and nearby galaxies (MRK 935, NGC 7810, and AGC 105092) outside the core group at $50^{\prime\prime}\times58^{\prime\prime}$ resolution, our deep FAST data return a much more complete picture. 

While the VLA primarily traces gas in galaxy discs and a tidal tail extending $\sim$130 kpc to the southwest of HCG 100 \citep{Jones2023,deMello2008}, FAST uncovers widespread, low-density \hi\ emission filling most of the field and connecting multiple galaxies. In particular, diffuse gas bridges link AGC 105092 with NGC 7810, and MRK 935 with both HCG 100 and NGC 7810. A comparison of \hi\ emission between FAST and VLA data under the same resolution is presented in \ref{app:convol} (Figure \ref{fig:vla_convolve}).
In addition, our FAST data detect six new \hi\ sources (HS), all associated with optical galaxies in the field (see Section \ref{subsec:hi_neighbors}).

The extended \hi\ envelope, which reaches a minimum integrated column density of $8.6\times10^{17}\,\rm cm^{-2}$, has a projected length of approximately 1 Mpc along its major axis (southeast-northwest direction, dashed black line in Figure \ref{fig:moment_maps}(a)) and a width of $\sim$0.65 Mpc along its minor axis. The \hi\ emission in the northeast and northwest probably extends even further, beyond the currently mapped sky region. 
The total \hi\ mass in the field is $(3.10\pm0.31) \times 10^{10}\, M_{\odot}$, which is significantly higher than the VLA result of $(1.45 \pm 0.17) \times 10^{10} M_\odot$ \citep{Jones2023}. The difference is due to both the missing of extended emissions by the VLA interferometry observations and the excellent sensitivity of FAST, which enables detections of very faint diffuse emissions. Excluding masses of the six new \hi\ sources, FAST data suggest that up to 45\% of \hi\ gas $ (\sim1.4 \times 10^{10}\, M_{\odot})$ resides in the large-scale environment instead of disks. 
The pink contours in Figure \ref{fig:moment_maps}(a) trace the distribution of gas with very low column density, $\nhi<3.4\times10^{18}\,\rm cm^{-2}$ (12$\sigma$). The total integrated flux of the very diffuse \hi\ is 1.45 $\pm$ 0.15 $\rm Jy\, km\, s^{-1}$, corresponding to $(2.04\pm0.21)\times 10^{9}\, M_{\odot}$, which accounts for only $\sim$5\% of the total gas mass.

\begin{figure*}[t]
\centering
\includegraphics[width=0.99\textwidth]{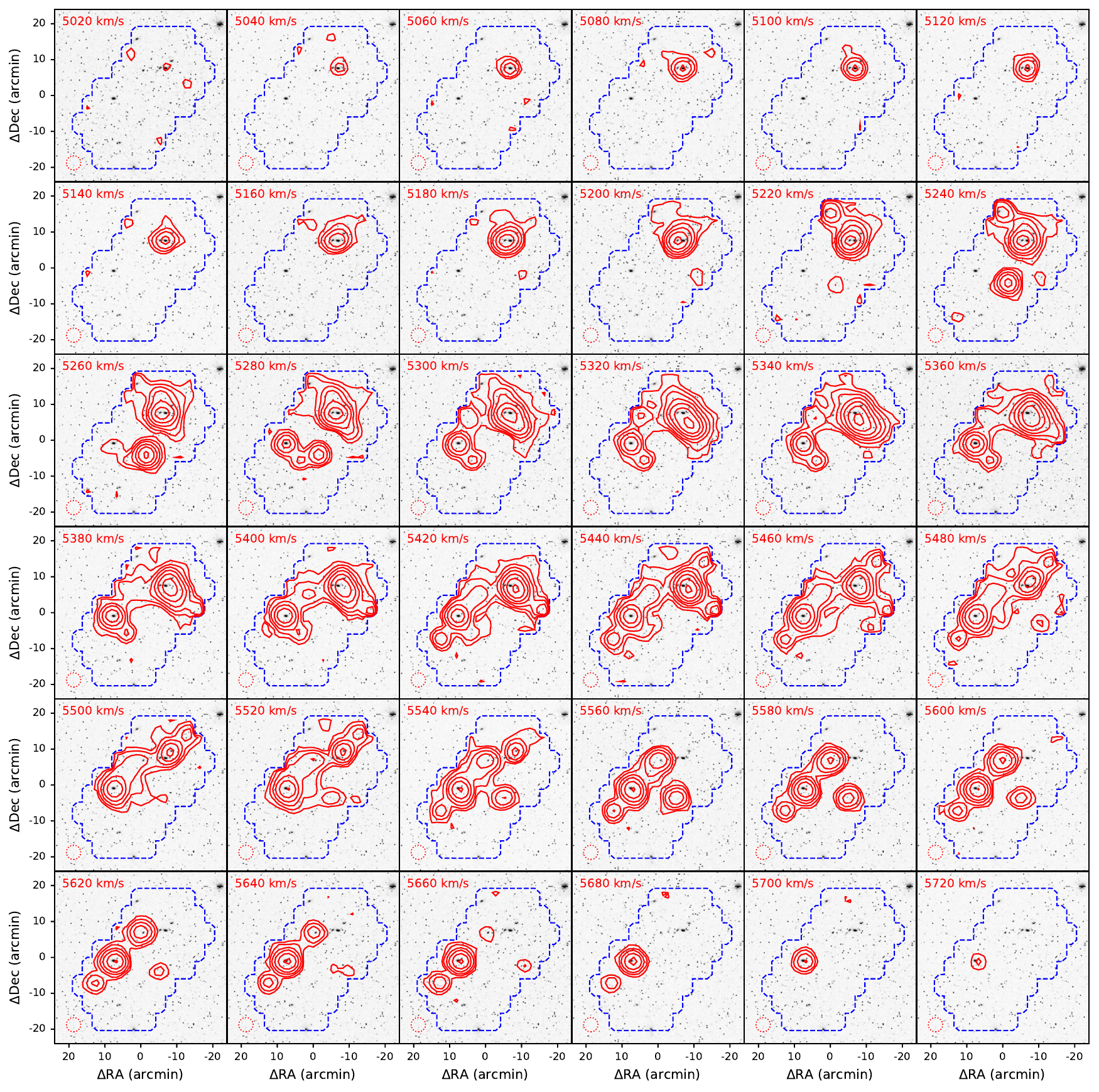}
\caption{\small FAST \hi\ channel maps (contours) overlaid on a DECaLS $r$-band image. The channel maps cover the velocity range of 5020$-$5720 ${\rm km\,s^{-1}}$ with a channel width of 20 ${\rm km\,s^{-1}}$. The contour levels start from $\nhi=1.4\times10^{17}\,\rm cm^{-2}$ (3$\sigma$) with an increment by a factor of 2. The red dashed circle in the bottom left corner represents the FAST beam after smoothing (4\arcmin). The blue dashed
lines outline the coverage of the FAST observations.
\label{fig:channel1}
}
\end{figure*}

In Figure \ref{fig:moment_maps}(b-c), we derived the velocities and velocity widths of the \hi\ emission from moment maps. From the velocity map of the \hi\ gas (Figure \ref{fig:moment_maps}(b)), we find that most of the extended \hi\ emission is distributed in the velocity range of 5200$-$5600 \kms. From channel maps (Figure \ref{fig:channel1}), we find that the diffuse \hi\ is detected in continuous channels. In particular, the bridge structure across NGC 7810, MRK 935, and HCG 100 is detected in the velocity range of 5300$-$5640 \kms\ (Figure \ref{fig:channel1}), while the VLA data suggest the emission of MRK 935 is centered on 5600 \kms\ within 5475$-$5667 \kms. We also identify large-scale positive or negative velocity gradients toward nearby converging directions, indicated by red or blue arrows in Figure \ref{fig:hi_flows_overlay}, suggestive of gas flows around the group.

Based on the high-resolution VLA data and channel maps (Figure \ref{fig:channel1}), the extended \hi\ emission around the group can be divided into two distinct velocity components: a 5300 \kms\ component (5200$-$5400 ${\rm km\,s^{-1}}$), tracing AGC 105092 and the long \hi\ tail attached to HCG 100 \citep{deMello2008}, and a 5500 \kms\ component (5400$-$5600 \kms), associated primarily with MRK 935 and NGC 7810. 
The compact group HCG 100 extends across both velocity bins.
Moment maps of the two components (Figure \ref{fig:moment_maps}(d-g)) indicate that the diffuse envelope arises predominantly from the contribution of the 5300 \kms\ component on the minor axis and of the 5500 \kms\ component on the major axis. 
Along the minor axis, the high-density region of the 5300 \kms\ component traces the tidal tail attached to HCG 100. 
Compared to VLA observations, the deeper FAST data highlight a more extended low-density ($4.6\times10^{17}\,\rm cm^{-2}$) gas that also connects HCG 100 with neighboring galaxies in the 5300 \kms\ component. Along the major axis, the diffuse \hi\ is dominated by the 5500 \kms\ component, with a minimum \hi\ column density of $4.6\times10^{17}\,\rm cm^{-2}$ and an elongated structure stretching over $\sim$0.9 Mpc. 

\subsection{Large-scale Velocity Gradients }\label{subsec:v_grad}
\begin{figure*}[!htbp]
\centering
\includegraphics[width=1.0\textwidth]{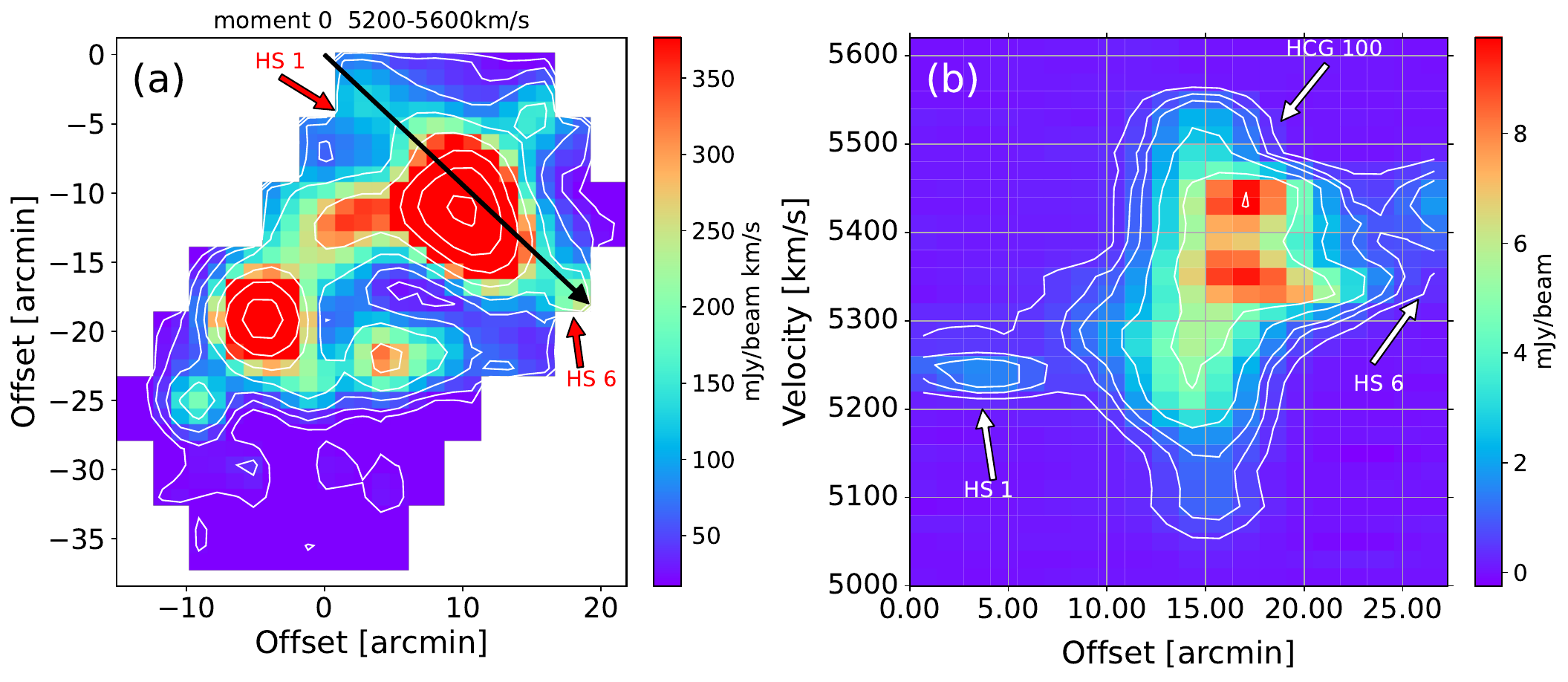}
\includegraphics[width=1.0\textwidth]{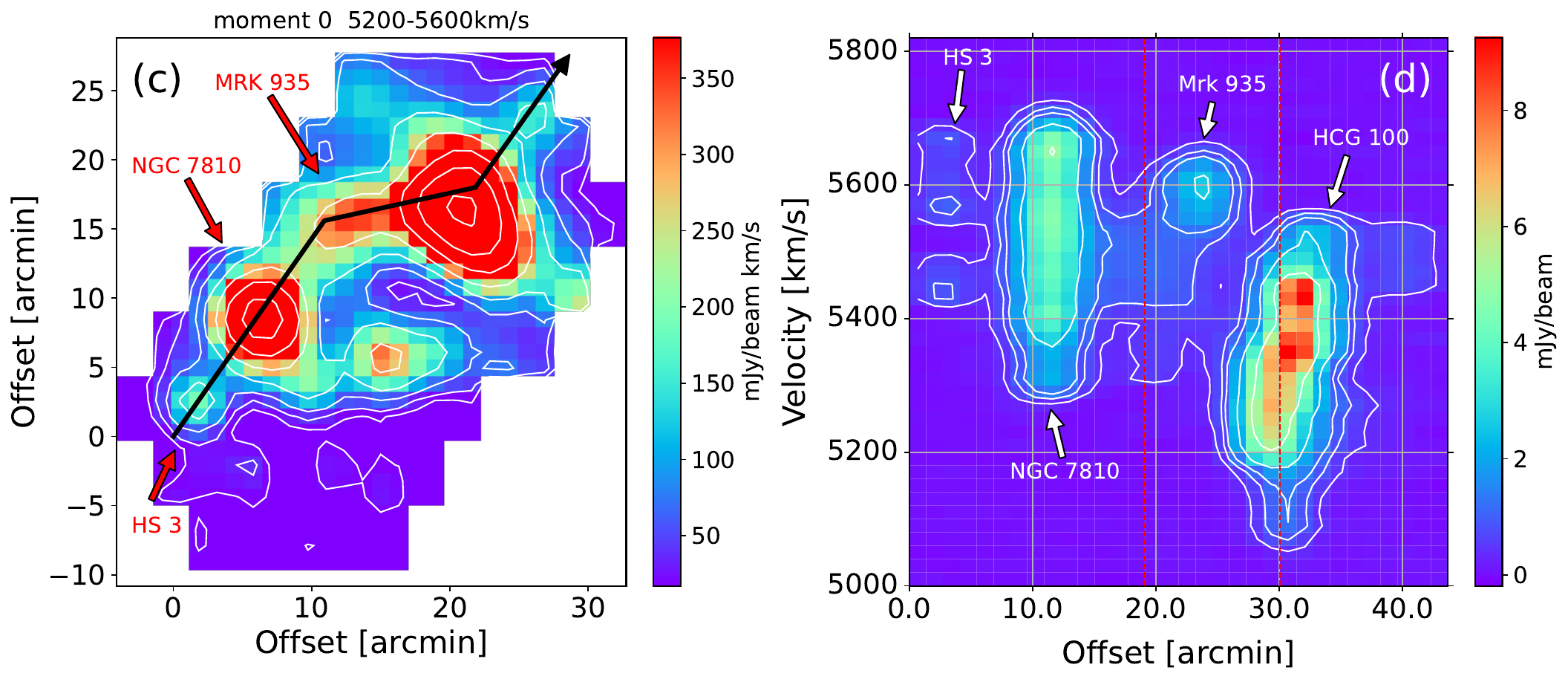}
\caption{
False color maps and position-velocity diagrams of the \hi\ gas. (a) Contours and false color map of the \hi\ emission integrated in the velocity range of 5200$-$5600 ${\rm km\,s^{-1}}$. The black line with an arrow represents the stripe adopted to extract the data and make the P–V diagram (panel (b)). The P-V diagrams start from offset = 0 and end at the arrow. The width of the stripe is 5$^{\prime}$. The contours have the base level of $\nhi=6.5\times10^{17}\,\rm cm^{-2}$ (3$\sigma$) and increase by a factor of 2. (b) P–V diagram following the direction illustrated in panel (a). (c) Contours and false color map of the \hi\ emission integrated in the velocity range of 5200$-$5600 ${\rm km\,s^{-1}}$. (d) P–V diagram following the direction illustrated in panel (c). The P–V diagram in panel (d) illustrates two turning points marked with red dashed lines. The contours in panels (c) and (d) start from 0.4 mJy beam$^{-1}$ ($\sim$6$\sigma$) with an increment of a factor of 2.
\label{fig:pv}
}
\end{figure*}

We further investigate the physical nature of the diffuse \hi\ by comparing moment maps (Figure \ref{fig:moment_maps}) and position-velocity (P-V) diagrams (Figure \ref{fig:pv}) of the two velocity components. In the 5300 \kms\ component (Figure \ref{fig:moment_maps}(d-f)), the low-density gas around HCG 100 is elongated towards the northeast and southwest directions, while MRK 935 is embedded in low-density \hi\ with $\nhi<3\times10^{18}\,\rm cm^{-2}$ with no high-density emission detected by the VLA. The \hi\ emissions of NGC 7810, HS 2, and AGC 105092 are connected by a bridge-like structure. In Figure \ref{fig:moment_maps}(e), the velocity field suggests a smooth gradient with line-of-sight velocities increasing from AGC 105092 to NGC 7810. The line width of the gas also increases when passing through HS 2 and NGC 7810 (Figure \ref{fig:moment_maps}(f)). 
These features may indicate that AGC 105092 and HS 2, and the \hi\ gas around them, are flowing towards NGC 7810. 

In Figure \ref{fig:pv}(a-b), we present the P-V diagram along the direction passing through the unresolved \hi\ sources HS 1 and HS 6 (black arrow). 
Along this line of sight, the gas shows a gradual velocity gradient of several tens \kms\ across a projected distance of roughly 0.6 Mpc centered on HCG 100, also clearly visible in Figure \ref{fig:moment_maps}(e). Most of the \hi\ emission in HCG 100 is detected in the velocity range 5050$-$5550 ${\rm km\,s^{-1}}$ (Figure \ref{fig:pv}(b), offset $\sim$15$^{\prime}$), showing asymmetric velocities toward the \hi\ tails attached to HCG 100. HS 1 and HS 6, which are likely associated with two gas-rich satellites of the compact group, are located at the start (offset $\sim$4$^{\prime}$) and end (offset $\sim$27$^{\prime}$) of the P-V path, with central velocities of 5230 ${\rm km\,s^{-1}}$ and 5430 ${\rm km\,s^{-1}}$, respectively. The P-V slice shows an S-shaped velocity distribution passing through HS 1, HCG 100, and HS 6 (Figure \ref{fig:pv}(b)), indicating a large dynamically-connected gaseous structure associated with the group environment.

Two physical scenarios can plausibly explain such a kinematical feature. One possibility is that the gas represents tidal debris produced by gravitational interactions. For example, we may expect a similar structure if HS 1 and HS 6 interacted with HCG 100 and moved apart in opposite directions, with their \hi\ gas stripped off in the process and potentially being accreted onto HCG 100.
Alternatively, similar S-shaped patterns are commonly observed in P-V diagrams taken along the kinematical major axis of rotating disks.
In this case, the \hi\ emission may trace a giant rotating gas structure with a diameter of $\sim$587 kpc (27$^{\prime}$).
Speculatively, such a rotating structure may be a signature of accretion of intergalactic gas \citep{Stewart2011} along a preferred plane from surrounding cosmic web filaments, with the group acting as a node where gas flows converge.

The \hi\ emission of the 5500 \kms\ component (Figure \ref{fig:moment_maps}(g)) shows an elongated shape along the direction going from HS 4 and HCG 100 to MRK 935, NGC 7810, HS 3, and HS 5. All these galaxies are connected with gas bridges, and the contours around MRK 935 show a tidal-like morphology, stretching between HCG 100 and NGC 7810. The velocity field (Figure \ref{fig:moment_maps}(h)) reveals that the \hi\ emission between HS 3, NGC 7810, and MRK 935 is smoothly connected, and the velocity increases towards MRK 935 and declines when approaching HCG 100. The \hi\ line widths of the gas around HS 3, NGC 7810, and MRK 935 is relatively large, suggestive of complex kinematics. In contrast, the \hi\ gas around HS 5 has a higher velocity and lower linewidths, which may indicate that this galaxy is more isolated and not strongly interacting yet with the group. 

\begin{deluxetable*}{lrrccccr}\setlength\tabcolsep{8pt}
\tablenum{1}
\centering
\tablewidth{0pt}
\caption{Neighboring Sources Detected by FAST \label{tab:neighbors}}
\tablehead{
\colhead{Source ID} & \colhead{R.A.} & \colhead{Decl.} & \colhead{S/N} & \colhead{$v_{\rm HI}$} & \colhead{log($\mhi$)}	& \colhead{log($M_{\star}$)$^a$} & \colhead{$v_{\rm opt}$}\\
\colhead{} & \colhead{(degree)} & \colhead{(degree)} & \colhead{} & \colhead{(km/s)} & \colhead{($M_{\odot}$)}	& \colhead{($M_{\odot}$)} & \colhead{(km/s)}}
\startdata
HS 1 (LEDA 1426263) & 0.4678 & 13.2461 & 25.1 & 5232 $\pm$ 20 & 8.67 $\pm$ 0.05 & - & -\\
HS 2 (LEDA 73211) & 0.5167 & 12.8913 & 24.7 & 5357 $\pm$ 20 & 8.50 $\pm$ 0.04 & 9.46 & 5370 $\pm$ 10$^b$\\
HS 3 (WISEA J000243.11+125138.8) & 0.6795  & 12.8607 & 19.1 & 5603 $\pm$ 20 & 8.61 $\pm$ 0.05 & 8.74 & 5610 $\pm$ 10$^b$\\
HS 4 (LEDA 1425381) &  0.2120 & 13.2072 & 23.8 & 5480 $\pm$ 20 & 8.44 $\pm$ 0.04 & - & -\\
HS 5 (WISEA J000125.99+125519.1) & 0.3583 & 12.9220 & 36.8 & 5565 $\pm$ 20 & 8.77 $\pm$ 0.05 & 9.01 & 5580 $\pm$ 10$^b$\\
HS 6 (WISEA J000040.01+125920.2) & 0.1673 & 12.9883 & 42.6 & 5412 $\pm$ 20 & 8.60 $\pm$ 0.04 & - & -\\
MRK 935 & 0.4385 & 13.1006 & 63.5 & 5561 $\pm$ 20 & 9.17 $\pm$ 0.04 & 9.81 & 5606 $\pm$ 30$^c$\\
NGC 7810 & 0.5799 & 12.9717 & 89.2 & 5515 $\pm$ 20 & 9.84 $\pm$ 0.04 & 10.52 & 5532 $\pm$ 29$^d$\\
AGC 105092 & 0.4197 & 12.9166 & 73.1 & 5257 $\pm$ 20 & 8.91 $\pm$ 0.04 & 7.90 & 5254 $\pm$ 20$^e$\\
\enddata
\tablecomments{ \\ 
$^a$ Stellar masses derived from SDSS photometry and adopted redshift at the distance of 77 Mpc.\\
$^b$ Optical velocity from DESI DR1 \citep{DESI2026}.\\
$^c$ Optical velocity from \cite{Petrosian2007}.\\
$^d$ Optical velocity from \cite{Huchra1983}.\\
$^e$ \hi\ velocity from \cite{Haynes2018}.
}
\end{deluxetable*}

\begin{figure*}[!htbp]
\centering
\includegraphics[width=0.99\textwidth]{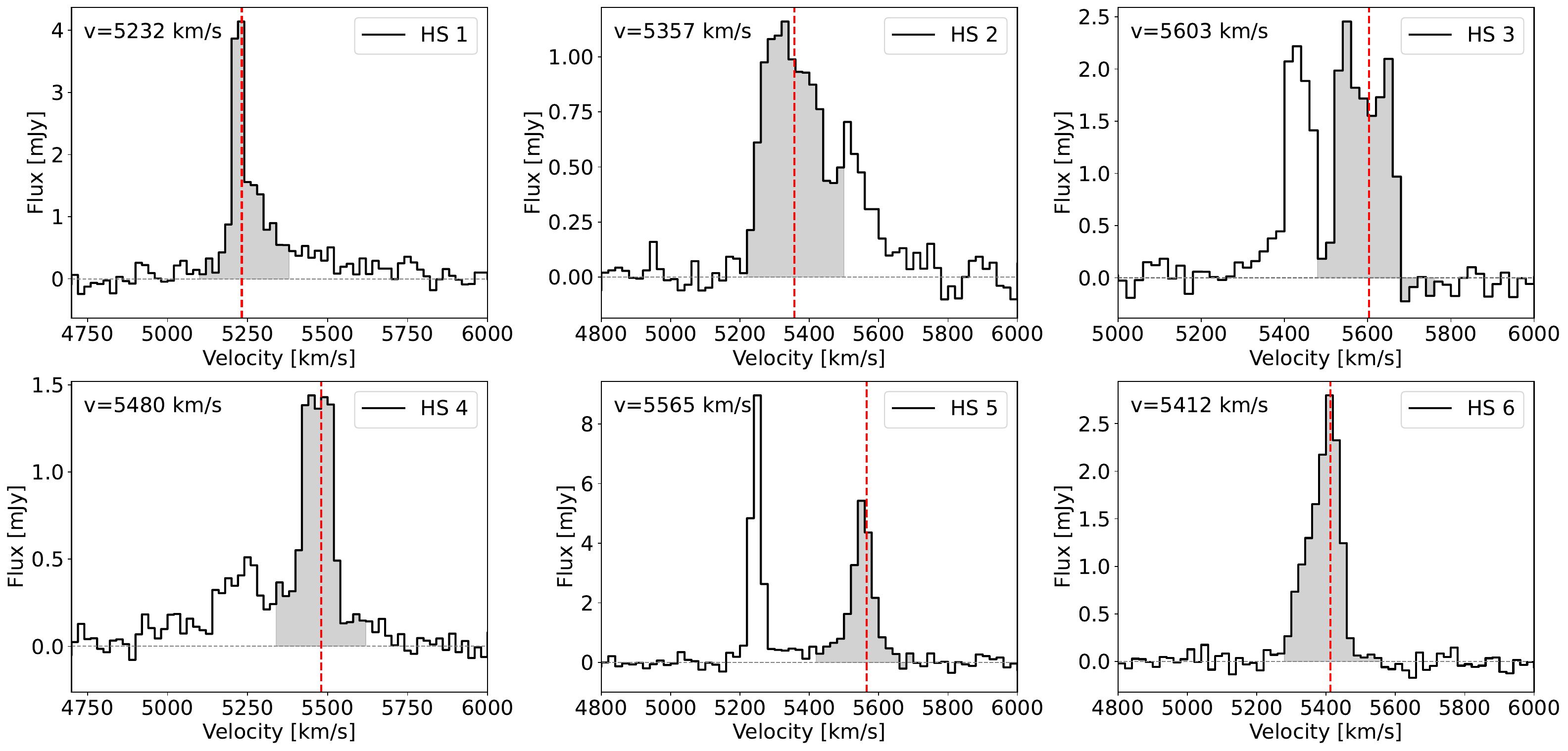}
\caption{
\hi\ spectra of six newly detected sources in the neighborhood of HCG 100. The red dashed lines show velocities of the \hi\ emission for each sources. The gray shaded regions represent velocity range used for the estimation of the corresponding \hi\ masses. The asymmetric wings and plateaus in the spectra are likely contributed by the diffuse intragroup \hi. The left peak in the spectrum of HS 5 is the emission from AGC 105092. HS 2, HS 3, and HS 5 have available optical spectra and redshifts from DESI, and their optical galaxies are confirmed by consistent radial velocities compared to \hi\ velocities. 
\label{fig:hs_spec}
}
\end{figure*}

We make P-V plots to inspect connections of \hi\ emission among HCG 100 and neighbouring galaxies. In Figure \ref{fig:pv}(c-d), the velocities along the direction passing through HS 3, NGC 7810, MRK 935, HCG 100, and HS 4 indicate that the low-density \hi\ is connected over a projected scale of 0.9 Mpc. The line-of-sight velocity of the gas steadily decreases while approaching HCG 100, which is suggestive of gas flowing onto the compact group. 
As before, the kinematics could be explained as gas tidally stripped because of multiple galaxy interactions: the tail/bridge features around MRK 935 may indicate that MRK 935 flew by NGC 7810 $\sim$1.2 Gyr ago, assuming a relative transverse velocity of 200 ${\rm km\,s^{-1}}$ for a distance of $\sim$254 kpc, and currently it is heading toward HCG 100. 
Another intriguing possibility is that the extended low-density, \hi\ gas represents a signature of cold accretion streams predicted by cosmological simulations.
In this scenario, pristine gas was accreted onto the gravitational well of the group, after which interactions among multiple galaxies redistribute the gas and reshape its large-scale flows.
During this process, several young blue systems (e.g., HS 2, HS 3, AGC 105092, and HS 5) may have formed within intergalactic star-forming regions triggered by tidal interactions and local \hi\ overdensities \citep{deMello2008,deMello2012}.

\subsection{Neighboring Galaxies around HCG 100}\label{subsec:hi_neighbors}
In the velocity range of 5200$-$5600 ${\rm km\,s^{-1}}$, we detected six new unresolved \hi\ sources in the neighborhood of HCG 100 (Table \ref{tab:neighbors}). Their \hi\ spectra are presented in Figure \ref{fig:hs_spec}. Each spectrum is extracted within a FAST beam centered on the source, and the velocity is in barycentric velocity with a resolution of 20 \kms. We examined the spectrum from DESI DR1 \citep{DESI2026} to confirm the associated optical galaxy for the \hi\ source. HS 2, HS 3, and HS 5 have available spectra and redshift measurements from DESI DR1, and the radial velocities from optical spectra are consistent with our \hi\ velocities. The other three sources have candidates for optical counterparts but lack spectroscopic redshifts. 

The extracted \hi\ spectra of these sources show asymmetric wings or plateaus (Figure \ref{fig:hs_spec}), which are likely contributed by the diffuse intragroup \hi\ emission. The distinct peak shown in the spectra of HS 5 is the emission from AGC 105092. To measure the \hi\ in galaxies, we first selected a velocity window of 300 \kms\ centered on the velocity of the \hi\ for each galaxy, which is initially indicated by either the optical redshift or the peak velocity. Then, we measured the line center within the spectral window, and the integrated flux and \hi\ mass were derived by integrating the 300 \kms\ velocity interval centered on the measured line center. Adopting a distance of 77 Mpc, their \hi\ masses are in the range of log($\mhi/M_{\odot}$) = 8.44$-$8.77 (Table \ref{tab:neighbors}).

We derived the stellar masses of HS 2, HS 3, and HS 5 (Table \ref{tab:neighbors}) by applying the SDSS photometry, the adopted redshift at the distance of 77 Mpc to the code FASTPP \citep{Kriek2009}. For this, we adopted the Chabrier initial mass function \citep{Chabrier2003} and Bruzual \& Charlot stellar population models \citep{Bruzual2003}. HS 2 is relatively massive with a stellar mass of $2.9\times10^{9}\, M_{\odot}$, while HS 3 and HS 5 have stellar masses of $5.5\times10^{8}\, M_{\odot}$ and $1.0\times10^{9}\, M_{\odot}$. Their stellar masses are consistent with those of dwarf galaxies seen in the local Universe \citep{McConnachie2012}.

\section{Discussion} \label{sec:discussions}

Overall, our study unveil the presence of diffuse neutral gas distributed between galaxies across Mpc scales, highlighting the importance of ultra-deep \hi\ observations for uncovering previously unseen components of the baryon cycle. 
The morphology and kinematics of this diffuse gas around HCG 100 can be explained either by tidal stripping during galaxy interactions or by accretion from the intergalactic medium (IGM).
HCG 100 and neighboring galaxies are associated with loose groups \citep{Ramella1994} and lie within a large-scale concentration of 17 compact groups extending over 93 $h^{-1}$ Mpc \citep{Palumbo1993}. 
This rich environment explains the large number of neighboring galaxies around HCG 100 within the velocity range of 5200$-$5600 ${\rm km\,s^{-1}}$, and may facilitate both galaxy interactions and gas accretion with large-scale coherent kinematics on scales of 0.5$-$1 Mpc.

\begin{figure}[!htbp]
\centering
\includegraphics[width=0.49\textwidth]{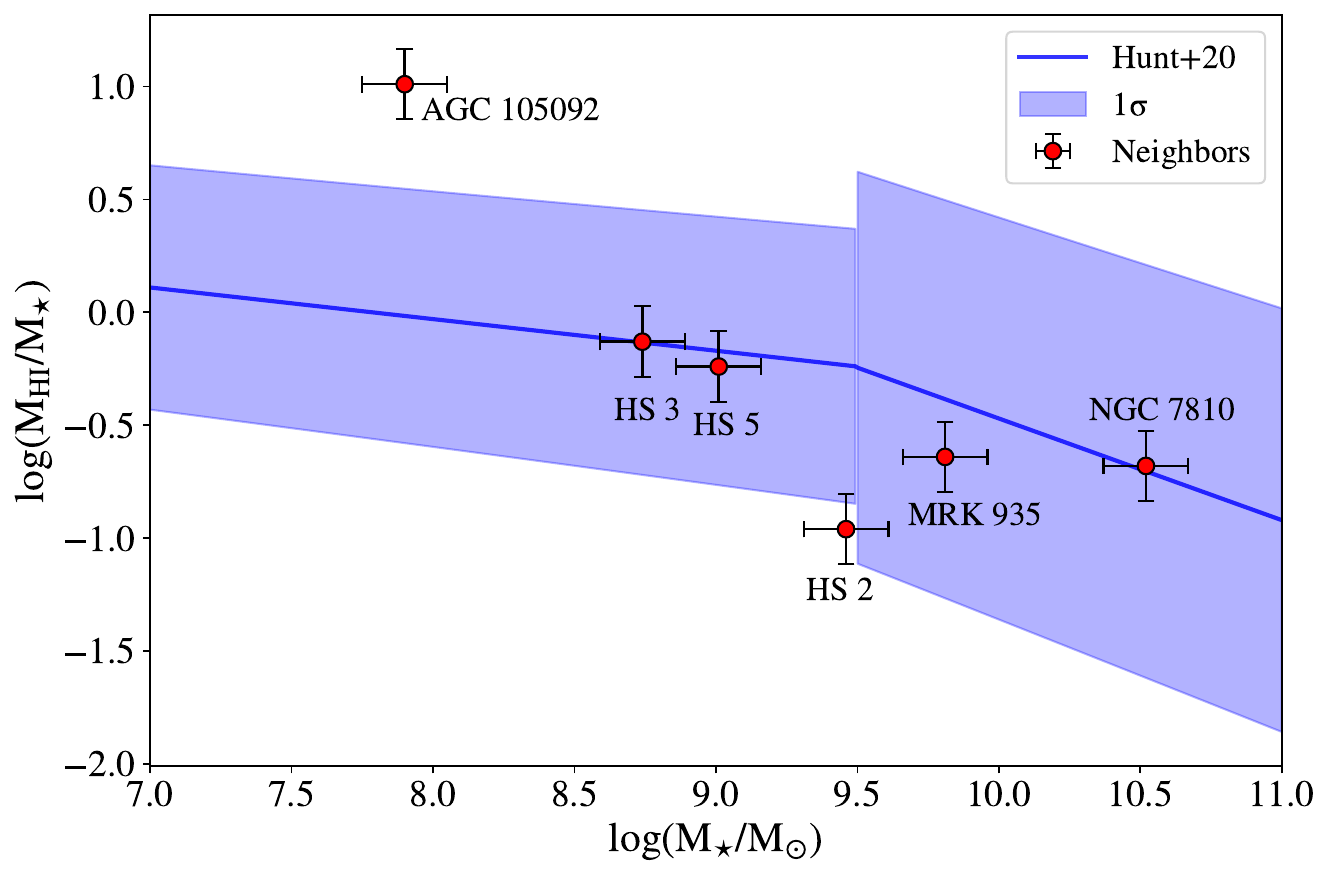}

\caption{
\hi\ fraction of neighboring galaxies as a function of stellar mass. The best-fit scaling relation of a sample of local star-forming galaxies is shown by a blue line \citep{Hunt2020}, and the shaded regions correspond to 1$\sigma$ deviations of the relation. 
\label{fig:scaling}
}
\end{figure}

If the diffuse gas arises from tidal stripping during past interactions between HCG 100 and nearby galaxies, low-mass systems would be expected to lose a larger fraction of their gas, potentially leading to \hi-deficient galaxies.
However, the \hi\ fractions of galaxies surrounding HCG 100 follow the stellar mass-\hi\ mass scaling relation for star-forming galaxies in the local Universe \citep[Figure~\ref{fig:scaling};][]{Hunt2020}, indicating no significant \hi\ deficiency.
This suggests that these neighboring systems were either initially very gas-rich prior to the interaction or they must have subsequently replenished their \hi\ reservoirs.

While cosmological simulations predict that cold gas streams can penetrate hot halos and rapidly feed star formation in galaxies with relatively low halo masses \citep[$M_\mathrm{halo}\lesssim10^{12}~M_\odot$;][]{Kere2005,Dekel2006,Dekel2009}, there is still little observational evidence of this process. 
Tentative signatures of this accretion mode have recently been reported at high redshift ($z>2$), where ionized and neutral gas streams extending over 
$>$100 kpc have been observed around galaxies \citep{Cantalupo2014,Martin2015,Emonts2023}. 

At low redshift, extended \hi\ structures larger than 100 kpc are commonly detected in and around interacting galaxies \citep{Heckman1982,Hibbard1999,Zhu2021}, groups \citep{Yun1994,Xu2022,Schneider1989,Michel-Dansac2010}, and clusters \citep{Koopmann2008}, although most of this diffuse gas can be attributed to tidal interactions. 
Recent observations of Stephan's Quintet have revealed diffuse \hi\ on scales of $\sim$0.6 Mpc that, similarly to the structure reported here for HCG 100, might also be explained by a scenario related to the cold gas accretion as an alternative to the tidal scenario \citep{Xu2022,Cheng2023}.

Both scenarios proposed here require a long dynamical timescale ($>$1 Gyr) to shape the \hi\ into its present configuration, raising the question of how such diffuse \hi\ can survive and remain neutral.
On the one hand, gas accreting from the IGM onto relatively massive halos is not expected to easily reach the star-forming disc through dense cold streams; instead, it should be shock-heated to the halo virial temperature and incorporated into a hot circumgalactic halo \citep{Dekel2009,Nelson2016}. 
Moreover, the intergalactic ultraviolet background is expected to ionize cold gas with column densities $\nhi\lesssim 2\times10^{19}\,\rm cm^{-2}$ on timescales of $\sim$500 Myr \citep{Corbelli1989,Borthakur2015}, suggesting that additional mechanisms are required to maintain the observed neutral gas. 
Clarifying the physical origin of this diffuse \hi\ therefore demands models capable of capturing the complex interplay between galaxies and their surrounding environments, following gas flows and phase transitions across the full range of scales, from the IGM down to galactic environments.

\section{Summary}\label{sec:Summary}
In conclusion, our ultra-deep FAST observations reveal a contiguous, low-column-density \hi\ structure extending to $\sim$1 Mpc around HCG 100. This \hi\ structure represents one of the most extended and lowest-density neutral gas structures detected in emission. It forms a coherent bridge connecting the group to nearby galaxies and shows large-scale coherent velocity gradients. We find that this structure contains $\sim$$1.4\times10^{10}\, M_{\odot}$ of diffuse \hi, indicating that up to 40$-$50\% of the total \hi\ resides outside the galaxies. The origin of such a large diffuse \hi\ structure can be explained either by large-scale tidal debris or gas accretion from the surrounding cosmic web. Our results demonstrate that massive reservoirs of diffuse neutral gas can persist on megaparsec scales around galaxy groups for extended periods, providing a previously unseen component of baryon cycling in dense environments. The long-lived Mpc-scale neutral gas structure provides a critical observational benchmark for theories of gas accretion and redistribution, requiring new generations of zoom-in cosmological simulations capable of resolving gas flows and phase transitions across from the IGM down to galactic environments.


\begin{acknowledgments}{\small
We thank the referee for careful reading and constructive comments that improved the paper. This work is supported by the National SKA Program of China No. 2025SKA0150103, National Natural Science Foundation of China under Nos. 12550002, 12133008, 12221003, 11890692. We acknowledge the science research grants from the China Manned Space Project with No. CMS-CSST-2021-A04 and No. CMS-CSST-2025-A10. Q.Y. and E.D.T.\ were supported by the European Research Council (ERC) under grant agreement no.\ 101040751. This work has used the data from the Five-hundred-meter Aperture Spherical radio Telescope (FAST).  FAST is a Chinese national mega-science facility, operated by the National Astronomical Observatories of Chinese Academy of Sciences (NAOC).

}
\end{acknowledgments}

%

\vspace{5mm}
\facilities{FAST}


\software{ Astropy \citep{astropy:2022},
Matplotlib \citep{Matplotlib2007}, 
NumPy \citep{Numpy2020},
SciPy \citep{Scipy2020},
FASTPP \citep{Kriek2009}.
          }


\appendix

\section{Sidelobe Correction}\label{app:sidelobe}

\begin{figure}[!htbp]
\centering
\includegraphics[width=0.49\textwidth]{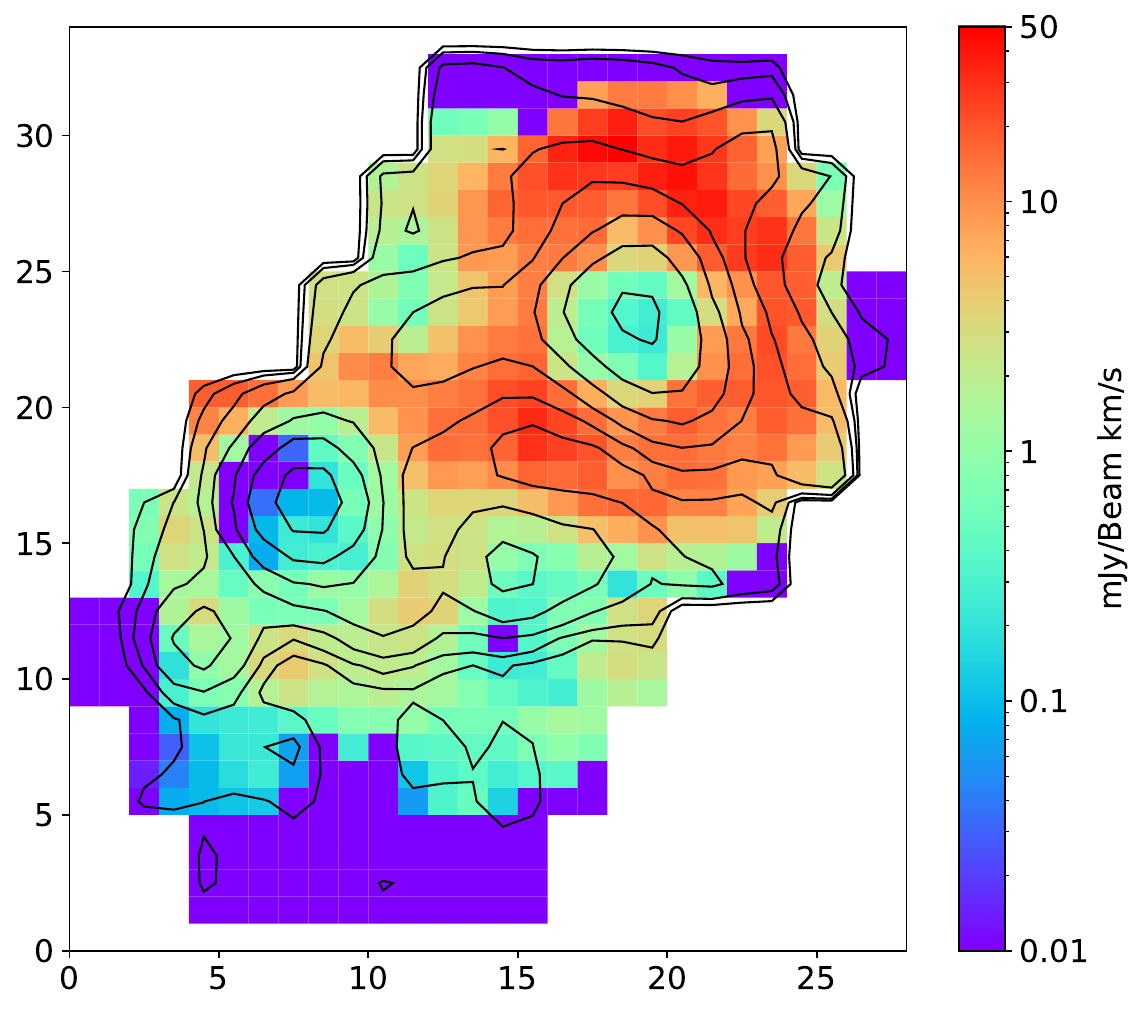}
\caption{Sidelobe contribution to the integrated \hi\ emission. False color image of the sidelobe contribution overlaid by the contour map of integrated \hi\ emission in the velocity range of 5200$-$5600 ${\rm km\,s^{-1}}$. The contours start from 17 mJy \kms\ $\text{beam}^{-1}$ and increase by a factor of 2.
\label{fig:sidelobe}
}
\end{figure}
 
To calculate the sidelobe contamination, we first selected the bright \hi\ sources that could make a non-negligible contribution to the sidelobe map. The first sidelobe response of the FAST 19-beam receiver can reach the level of $\sim$ 1\% of the peak response \citep{Jiang2020, Xu2022}. Therefore, a source would produce a sidelobe contribution comparable to the 1$\sigma$ level of the final integrated map only if its peak integrated flux is larger than $\sigma_{\rm M0}/0.01$. Since the lowest contour in Figure \ref{fig:sidelobe} corresponds to $3\sigma_{\rm M0}=17$ \mjb\,\kms, this gives a threshold of $\sim$570 \mjb\,\kms. We therefore included only sources with FAST-resolution peak integrated fluxes above the threshold in the sidelobe correction. The four sources HCG 100, NGC 7810, Mrk 935, and AGC 105092 satisfy this criterion, while the other \hi\ detections in the field are below this limit and their expected sidelobe contributions are therefore below the noise level of the integrated map.

These four bright sources were detected with high-resolution VLA observations, so we used the VLA data to derive the spatial distributions of the \hi\ emission for each source as the approximation for the ``truth map''. To avoid introducing noise or unrelated low-level emission into the sidelobe model, the source masks were defined using circles centered on sources in the flux-scaled VLA map smoothed to the FAST resolution. Pixels outside masks were excluded from the truth maps used for the sidelobe calculation. To account for the missing short-spacing flux in the interferometric data, we smoothed the VLA data to the FAST resolution and compared it with the corresponding FAST emission. By comparing the peak flux measured in the raw FAST map and that in the smoothed VLA map, we derived a flux-scaling ratio for each bright source. The scale factors are 1.28, 1.18, 1.08, and 1.48 for HCG 100, NGC 7810, Mrk 935, and AGC 105092, respectively. These flux-scaling ratios were then applied to the original VLA maps of the selected source regions to construct the high-resolution truth maps.

To derive the sidelobe contribution, we use the continuum observations of individual beams of the 19-beam receiver \citep{Jiang2020} as measured beam maps. For each beam, the ``main beam'' is approximated by a 2-D Gaussian with the FWHM equal to the HPBW. Using the high-resolution VLA truth maps, the measured FAST beam maps and the corresponding Gaussian main beams were re-sampled to the VLA grid, then the sidelobe contribution for each beam is defined by the difference between the measured FAST beam and the main beam. The sidelobe maps of the four bright sources were derived separately. After deriving the sidelobe maps of the four sources for each beam, the final sidelobe maps are assigned and re-gridded to the FAST grid based on the pointing configuration. The configuration of each pointing of FAST observations is listed in Table \ref{tab:obs}.

Figure \ref{fig:sidelobe} shows the derived total sidelobe contribution from the four bright sources, with contours showing the integrated \hi\ emission over 5200$-$5600 \kms before sidelobe correction. The modeled sidelobe contamination is spatially concentrated in ring-line structures around the peaks of the bright \hi\ sources, especially HCG 100, as expected from the FAST beam response. This morphology is clearly different from the extended diffuse \hi\ structures detected between the galaxies. Quantitatively, the mean modeled sidelobe contribution is 5.5 \mjb\,\kms, with a maximum of $\sim$ 50 \mjb\,\kms near the bright-source peaks. In contrast, the diffuse emission between the bright sources is typically detected between the third and fourth contours, corresponding to 68$-$136 \mjb\,\kms. At the locations of the diffuse \hi\ structures, the modeled sidelobe contribution is much lower, typically at the level of $\sim$0.1\% of the bright-source peak response and more than an order of magnitude below the detected diffuse emission. The mismatch between the spatial distribution of significant sidelobe contamination and that of the diffuse \hi\ therefore provides an important check that the extended low-column-density emission is not produced by residual sidelobes.

Because the available FAST beam maps were measured in continuum, we assume that the beam response does not vary significantly over the narrow frequency range covered by the \hi\ emission \citep{Jiang2020}. The velocity dependence of the sidelobe correction is instead determined by the \hi\ spectra of the bright sources. At the FAST resolution of $\sim$2$\farcm$9, the dominant sidelobe contribution is produced mainly by the compact bright peaks of the \hi\ sources. For the purpose of estimating this contribution, each bright source can therefore be described, to first order, by its position, integrated flux distribution, and spectrum. The first two quantities are provided by the source-specific VLA-based moment-0 truth maps, while the third is included through the source-specific spectral scaling applied to each velocity channel.

Then, we estimate the sidelobe contribution separately for each dominant bright \hi\ source, namely HCG 100, NGC 7810, Mrk 935, and AGC 105092. For each source ($s$), we construct a source-specific integrated sidelobe map, $M_{{\rm side},s}(x,y)$ in \mjb\,\kms, from the scaled VLA truth map and the measured FAST beam response. The contribution of this source to a velocity channel ($v$) is then scaled according to its own \hi\ spectrum,
\begin{equation}
    S_{\rm side}(x,y,v)=\sum_s C_s(v),M_{{\rm side},s}(x,y),
\end{equation}
where $C_s(v)=S_s(v)/S_{s,5200-5600}$, $S_s(v)$ is the flux density (in mJy) of source in the given channel, and ($S_{s,5200-5600}$) is its integrated flux (in mJy \kms) over 5200$-$5600 \kms. The summed source-specific sidelobe contribution is then subtracted from each channel of the FAST cube to obtain the sidelobe-corrected data cube.

\section{Comparison of \hi\ Emission with Convolved VLA Data}\label{app:convol}
\begin{figure*}[!h]
\centering
\includegraphics[width=\textwidth]{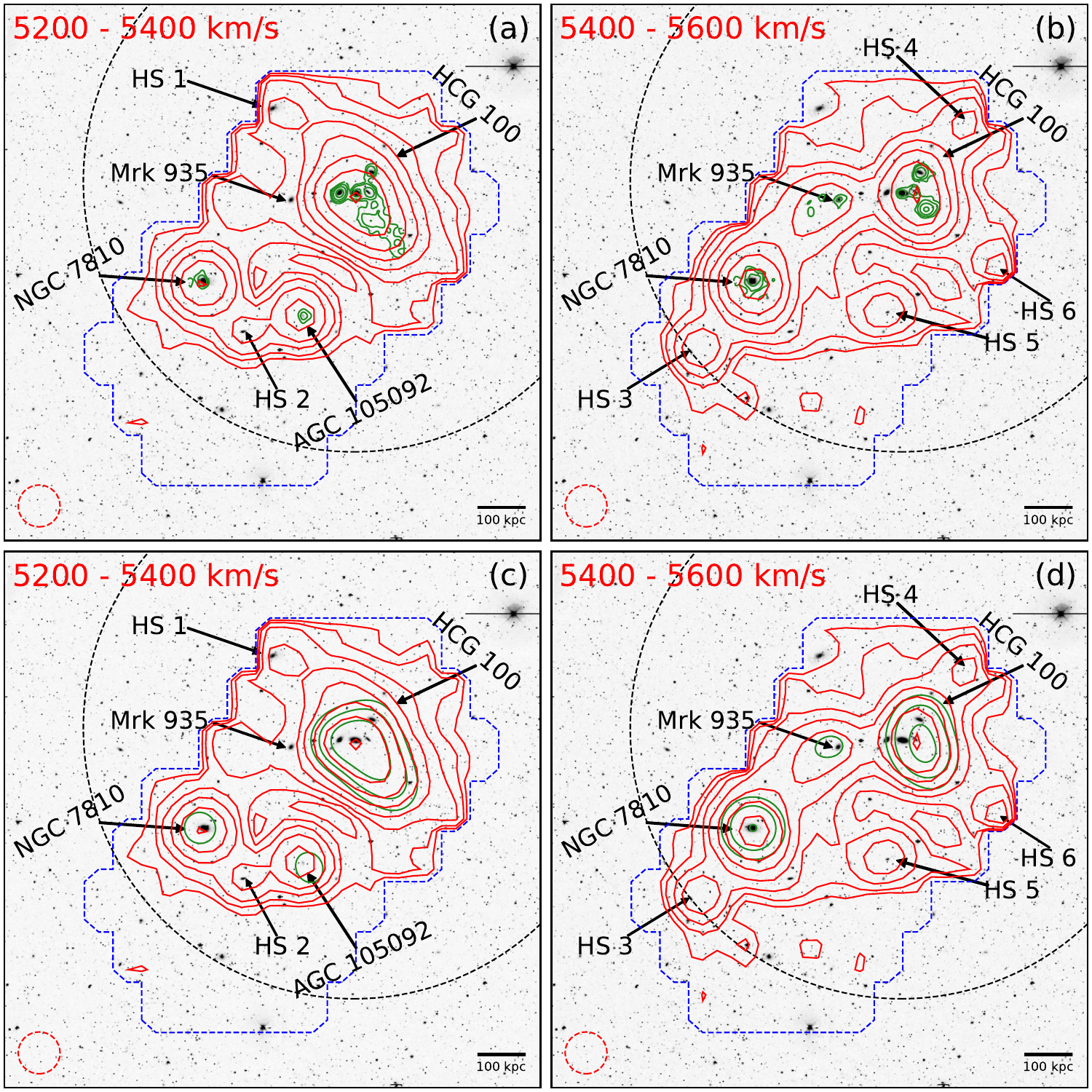}
\caption{
Comparison of \hi\ emission with convolved VLA data. Panels (a) and (b): contours of the integrated \hi\ emission in the velocity range of 5200$-$5600 ${\rm km\,s^{-1}}$ overlaid on the DECaLS $r$-band image. The FAST contours in red have the base level of $\nhi=4.6\times10^{17}\,\rm cm^{-2}$ (3$\sigma$) and increase by a factor of 2. The green contours from VLA observations start from $\nhi=5.0\times10^{19}\,\rm cm^{-2}$ (3$\sigma$) and increase by a factor of 2. The red dashed circle in the bottom left corner represents the FAST beam after smoothing (4$^{\prime}$). The blue dashed lines outline the boundary of the FAST observations, while the black dashed circles represent the footprint of the VLA observations. Panels (c) and (d): the same contour maps as in panels (a) and (b) with the VLA data convolved to the same beam shape as FAST observations. After convolution, the green contours from VLA observations start from $\nhi=1.7\times10^{19}\,\rm cm^{-2}$ (3$\sigma$) and increase by a factor of 2.
\label{fig:vla_convolve}
}
\end{figure*}

In order to investigate the possible beam smearing effects of FAST observations on bright \hi\ emission, we convolved the VLA data to the same beam of FAST for a comparison. The original data cube from VLA has an angular resolution of $50^{\prime\prime}\times58^{\prime\prime}$, which was convolved into the same resolution as the FAST data after smoothing and sidelobe correction ($4^{\prime}\times4^{\prime}$). As shown in  Figure \ref{fig:vla_convolve}, the contour maps from VLA after convolution are consistent with those of the high-density \hi\ detected by FAST. Beam smearing effects can not justify the extended diffuse structure detected in the FAST data cube.

\begin{deluxetable*}{lrrccccr}
\tablenum{A1}
\centering
\tablewidth{0pt}
\caption{FAST observations \label{tab:obs}}
\tablehead{
\colhead{Observing ID} & \colhead{R.A.} & \colhead{Decl.} & \colhead{Date} & \colhead{$t_{\rm total}$} & \colhead{$f_{\rm central}$}	& \colhead{On-OFF} & \colhead{$t_{\rm on}$}\\
\colhead{} & \colhead{} & \colhead{} & \colhead{} & \colhead{(s)} & \colhead{(MHz)}	& \colhead{Cycles} & \colhead{(s)} }
\startdata
HCG100\_p1 & 00:01:13.76 & +13:07:44.44 & 2023-09-08 & 4260 & 1395.56 & 6 & 1800 \\
HCG100\_p2 & 00:01:19.36 & +13:07:44.44 & 2023-09-09 & 4260 & 1395.56 & 6 & 1800 \\
HCG100\_p3 & 00:01:24.96 & +13:07:44.44 & 2023-09-12 & 4260 & 1395.56 & 6 & 1800 \\
HCG100\_p4 & 00:01:30.56 & +13:07:44.44 & 2023-09-14 & 4260 & 1395.56 & 6 & 1800 \\
HCG100\_p5 & 00:01:13.76 & +13:06:32.44 & 2023-09-15 & 4260 & 1395.56 & 6 & 1800 \\
HCG100\_p6 & 00:01:19.36 & +13:06:32.44 & 2023-09-16 & 4260 & 1395.56 & 6 & 1800 \\
HCG100\_p7 & 00:01:24.96 & +13:06:32.44 & 2023-09-23 & 4260 & 1395.56 & 6 & 1800 \\
HCG100\_p8 & 00:01:30.56 & +13:06:32.44 & 2023-09-26 & 4260 & 1395.56 & 6 & 1800 \\
HCG100\_p9 & 00:01:13.76 & +13:05:20.44 & 2023-10-05 & 4260 & 1395.56 & 6 & 1800 \\
HCG100\_p10 & 00:01:19.36 & +13:05:20.44 & 2023-10-18 & 4260 & 1395.56 & 6 & 1800 \\
HCG100\_p11 & 00:01:24.96 & +13:05:20.44 & 2023-10-24 & 4260 & 1395.56 & 6 & 1800 \\
HCG100\_p12 & 00:01:30.56 & +13:05:20.44 & 2023-10-29 & 4260 & 1395.56 & 6 & 1800 \\
HCG100\_p13 & 00:01:13.76 & +13:04:08.44 & 2023-11-14 & 4260 & 1395.56 & 6 & 1800 \\
HCG100\_p14 & 00:01:19.36 & +13:04:08.44 & 2023-11-19 & 4260 & 1395.56 & 6 & 1800 \\
HCG100\_p15 & 00:01:24.96 & +13:04:08.44 & 2023-11-23 & 4260 & 1395.56 & 6 & 1800 \\
HCG100\_p16 & 00:01:30.56 & +13:04:08.44 & 2023-11-29 & 4260 & 1395.56 & 6 & 1800 \\
HCG100\_p17 & 00:01:59.76 & +12:52:57.04 & 2024-08-31 & 1860 & 1395.56 & 3 & 780 \\
HCG100\_p18 & 00:02:05.36 & +12:52:57.04 & 2024-08-31 & 1860 & 1395.56 & 3 & 780 \\
HCG100\_p19 & 00:02:10.96 & +12:52:57.04 & 2024-09-14 & 1860 & 1395.56 & 3 & 780 \\
HCG100\_p20 & 00:02:16.56 & +12:52:57.04 & 2024-09-14 & 1860 & 1395.56 & 3 & 780 \\
HCG100\_p21 & 00:01:59.76 & +12:51:45.04 & 2024-09-14 & 1860 & 1395.56 & 3 & 780 \\
HCG100\_p22 & 00:02:05.36 & +12:51:45.04 & 2024-09-14 & 1860 & 1395.56 & 3 & 780 \\
HCG100\_p23 & 00:02:10.96 & +12:51:45.04 & 2024-09-25 & 1860 & 1395.56 & 3 & 780 \\
HCG100\_p24 & 00:02:16.56 & +12:51:45.04 & 2024-09-25 & 1860 & 1395.56 & 3 & 780 \\
HCG100\_p25 & 00:01:59.76 & +12:50:33.04 & 2024-09-29 & 1860 & 1395.56 & 3 & 780 \\
HCG100\_p26 & 00:02:05.36 & +12:50:33.04 & 2024-09-29 & 1860 & 1395.56 & 3 & 780 \\
HCG100\_p27 & 00:02:10.96 & +12:50:33.04 & 2024-10-05 & 1860 & 1395.56 & 3 & 780 \\
HCG100\_p28 & 00:02:16.56 & +12:50:33.04 & 2024-10-06 & 1860 & 1395.56 & 3 & 780 \\
HCG100\_p29 & 00:01:59.76 & +12:49:21.04 & 2024-10-10 & 1860 & 1395.56 & 3 & 780 \\
HCG100\_p30 & 00:02:05.36 & +12:49:21.04 & 2024-10-10 & 1860 & 1395.56 & 3 & 780 \\
HCG100\_p31 & 00:02:10.96 & +12:49:21.04 & 2024-10-12 & 1860 & 1395.56 & 3 & 780 \\
HCG100\_p32 & 00:02:16.56 & +12:49:21.04 & 2024-10-12 & 1860 & 1395.56 & 3 & 780 
\enddata
\tablecomments{ 
}
\end{deluxetable*}



\bibliography{ref}{}
\bibliographystyle{aasjournal}



\end{CJK*}
\end{document}